\documentclass[preprint,aps,prl,nofootinbib,superscriptaddress]{revtex4-1}
\usepackage[utf8]{inputenc}

\usepackage{amsmath}    
\usepackage{graphicx}   
\usepackage{verbatim}   
\usepackage{color}      
\usepackage{epstopdf}  
\usepackage{upgreek} 
\usepackage[hidelinks]{hyperref}   
\usepackage{textcomp} 
\usepackage{siunitx}
\usepackage{dcolumn}

\newcommand{\figref}[1]{Fig.~\ref{#1}}
\newcommand{\tabref}[1]{Table~\ref{#1}}
\newcommand{\verref}[1]{eq.~\eqref{#1}}

\begin{document}

\title{Extraction of Dzyaloshinksii-Moriya interaction from propagating spin waves validated}
\author{Juriaan Lucassen}
\email{j.lucassen@tue.nl}
\affiliation{Department of Applied Physics, Eindhoven University of Technology, P.O. Box 513, 5600 MB Eindhoven, the Netherlands}
\author{Casper F. Schippers}
\affiliation{Department of Applied Physics, Eindhoven University of Technology, P.O. Box 513, 5600 MB Eindhoven, the Netherlands}
\author{Marcel A. Verheijen}
\affiliation{Department of Applied Physics, Eindhoven University of Technology, P.O. Box 513, 5600 MB Eindhoven, the Netherlands}
\affiliation{Eurofins Materials Science BV, High Tech Campus 11, 5656 AE Eindhoven, The Netherlands}
\author{Patrizia Fritsch}
\affiliation{IFW-Dresden, Institute for Solid State Research, Helmholtzstraße 20, 01069 Dresden, Germany}
\author{Erik Jan Geluk}
\author{Beatriz Barcones}
\affiliation{NanoLab@TU/e, Eindhoven University of Technology, P.O. Box 513, 5600 MB Eindhoven, the Netherlands}
\author{Rembert A. Duine}
\affiliation{Department of Applied Physics, Eindhoven University of Technology, P.O. Box 513, 5600 MB Eindhoven, the Netherlands}
\affiliation{Institute for Theoretical Physics, Utrecht University, Princetonplein 5, 3584 CC Utrecht, the Netherlands}
\author{Sabine Wurmehl}
\affiliation{IFW-Dresden, Institute for Solid State Research, Helmholtzstraße 20, 01069 Dresden, Germany}
\affiliation{Institute of Solid State and Materials Physics, TU Dresden, 01062 Dresden, Germany}
\author{Henk J.M. Swagten}
\affiliation{Department of Applied Physics, Eindhoven University of Technology, P.O. Box 513, 5600 MB Eindhoven, the Netherlands}
\author{Bert Koopmans}
\affiliation{Department of Applied Physics, Eindhoven University of Technology, P.O. Box 513, 5600 MB Eindhoven, the Netherlands}
\author{Reinoud Lavrijsen}
\affiliation{Department of Applied Physics, Eindhoven University of Technology, P.O. Box 513, 5600 MB Eindhoven, the Netherlands}

\date{\today}

\begin{abstract}
The interfacial Dzyaloshinskii–Moriya interaction (iDMI) is of great interest in thin-film magnetism because of its ability to stabilize chiral spin textures. It can be quantified by investigating the frequency non-reciprocity of oppositely propagating spin waves. However, as the iDMI is an interface interaction the relative effect reduces when the films become thicker making quantification more difficult. Here, we utilize all-electrical Propagating Spin Wave Spectroscopy (PSWS) to disentangle multiple contributions to spin wave frequency non-reciprocity to determine the iDMI. This is done by investigating non-reciprocities across a wide range of magnetic layer thicknesses (from $4$ to $26$~\si{nm}) in Pt/Co/Ir, Pt/Co/Pt, and Ir/Co/Pt stacks. We find the expected sign change in the iDMI when inverting the stack order, and a negligible iDMI for the symmetric Pt/Co/Pt. We additionally extract a difference in surface anisotropies and find a large contribution due to the formation of different crystalline phases of the Co, which is corroborated using nuclear magnetic resonance and high-resolution transmission-electron-microscopy measurements. 
These insights will open up new avenues to investigate, quantify and disentangle the fundamental mechanisms governing the iDMI, and pave a way towards engineered large spin-wave non-reciprocities for magnonic applications.

%
%
%
\end{abstract}

\maketitle
Within magnetism, the interfacial Dzyaloshinskii–Moriya interaction (iDMI) has gained enormous interest in recent years. It is an antisymmetric exchange interaction generated at symmetry-breaking interfaces with high spin-orbit coupling~\cite{PhysRevB.78.140403,PhysRevLett.87.037203}, which can stabilize non-collinear spin textures such as magnetic skyrmions.~\cite{Bode2007,Emori2013,Ryu2013,PhysRevLett.87.037203,Fert2013,PhysRevB.78.140403} Because of its importance in the field of non-collinear spin textures, it is vital to get a fundamental understanding of this interaction. For this, one requires methods that are able to accurately determine the iDMI. There are several techniques that have the ability to do this, and they can be split up into two major categories. On the one hand, there are domain-wall based methods, which look at a magnetic domain-wall texture and/or its motion under the influence of symmetry-breaking magnetic fields.~\cite{PhysRevB.88.214401,Ryu2013,PhysRevB.93.144409,doi:10.1021/acs.nanolett.6b01593} 
The second branch contains spin-wave based methods which rely on the iDMI-induced frequency difference between oppositely propagating spin waves, and is commonly measured using the Brillouin Light Scattering (BLS) technique.~\cite{PhysRevB.88.184404,0953-8984-25-15-156001,Lee2016,Nembach2015,Cho2015,PhysRevLett.114.047201} Spin-wave based methods carry two major advantages: they do not require knowledge of the exchange interaction and they probe sample-averaged properties.~\cite{PhysRevB.94.104431}

When it comes to quantifying iDMI, BLS is limited with respect to the frequency difference that can be measured, and is therefore only suited to reliably measure the iDMI in thin film ($\sim 1\operatorname{-}2$~\si{nm}) systems with a large iDMI to generate enough non-reciprocity.~\cite{PhysRevB.88.184404,0953-8984-25-15-156001,Lee2016,Nembach2015,Cho2015,PhysRevLett.114.047201} Recently, all-electrical Propagating Spin Wave Spectroscopy (PSWS)~\cite{doppler} has been proposed as an alternative for probing this frequency difference.~\cite{Lee2016,PhysRevB.93.235131} As this technique is more sensitive to small frequency differences (few~\si{MHz} compared to tens to hundreds of MHz for BLS~\cite{PhysRevB.93.054430,10.3389/fphy.2015.00035}), the lower bound of iDMI that can be quantified is significantly improved and allows for the non-reciprocity to be investigated in thicker films ($\sim 20$~\si{nm}), well beyond the thickness limit of BLS. For these thicker films, however, additional effects can play a role; for example, spin-wave localization in combination with a difference in interfacial anisotropy of the top and bottom interface can also lead to frequency differences between oppositely propagating spin waves.~\cite{PhysRevB.93.054430} 

In this Letter, we therefore systematically untangle different contributions to the spin-wave frequency non-reciprocity utilizing PSWS to extract the iDMI. By investigating the non-reciprocity as a function of Co layer thickness $t$ for Pt/Co/Ir, Pt/Co/Pt and Ir/Co/Pt systems we isolate the iDMI from other contributions to the non-reciprocity.~\cite{PhysRevB.93.054430} For Pt/Co/Ir and Ir/Co/Pt we expect to find large but inverted DMI values, whilst the effective DMI for the symetric Pt/Co/Pt should be very small because the global symmetry is no longer broken.~\cite{PhysRevB.78.140403,PhysRevLett.87.037203,doi:10.1021/acs.nanolett.6b01593} This is indeed what we find for thin Co, where we also find the expected $1/t$ dependence of the non-reciprocity due to the interfacial nature of the iDMI. However, for thicker layers the non-reciprocities are dramatically enhanced by a hitherto unconsidered effect; a change in the crystal phase of Co above a thickness of $\sim 10$~\si{nm}. Nevertheless, also in this regime the iDMI can be reliable extracted, further substantiating the powerful nature of PSWS to extract the iDMI over a large thickness range.

\begin{figure}
\centering
\includegraphics{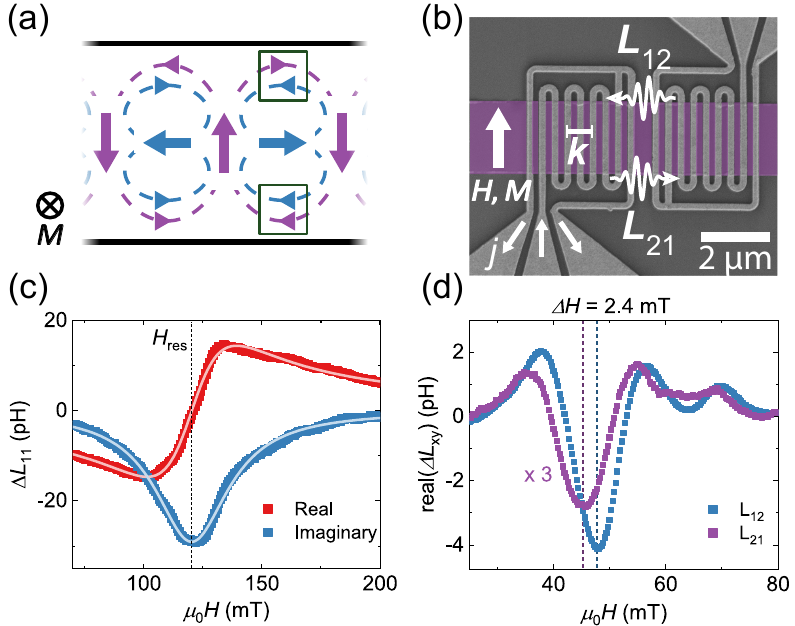}
\caption{\label{fig:figure1}(a) Sketch of a sideview of a thin magnetic layer where the solid arrows indicate the dynamic components of the magnetization for a clockwise spin wave. The dashed lines are the resulting stray fields where the two boxes highlight the additive and destructive interference of the resulting stray fields at the bottom and top of the layer, respectively, which leads to spin-wave localization. Adapted from Ref.~\onlinecite{PhysRevB.93.054430}. (b) SEM micrograph of a fabricated device with $k=8.5$~\si{\mu m^{-1}}. The magnetic strip is marked with a false color. We indicate the direction of the applied magnetic field $H$, current flow direction $j$, and the spin-wave flow direction given by the mutual inductions of the antennas ($L_\mathrm{xy}$) and sign of $k$. (c) Self-induction $\Delta L_\mathrm{11}$ as a function of applied magnetic field $H$ measured at $15$~GHz on Pt/Co($15$)/Ir with $k=7$~\si{\mu m^{-1}}. The dashed line indicates the resonance field $H_\mathrm{res}$ extracted from a fit of the resonance (solid lines). (d) Real part of the mutual-induction $\Delta L_\mathrm{xy}$ (with a rescaled $L_\mathrm{21}$) as a function of applied magnetic field $H$ measured at $11$~GHz on Pt/Co($12.3$)/Ir with $k=8.5$~\si{\mu m^{-1}}. The dashed lines demonstrate a measured peak shift $\Delta H$ of $\sim 2.4$~mT.}
\end{figure}

Before we describe the main results of this Letter, we first demonstrate how spin-wave localization can also lead to a frequency non-reciprocity. 
This localization is a consequence of an asymmetry in the dynamic dipolar fields of a spin wave, which is illustrated in~\figref{fig:figure1}a. In this figure we show the dynamic components of the magnetization of a clockwise (CW) spin wave including the resulting dipolar fields. As indicated with the green boxes, these dipolar fields add up constructively at the bottom of the film, and destructively near the top of the film. This asymmetry will localize the spin-wave on either the top or bottom interface, depending on the thickness of the magnetic film.~\cite{PhysRevB.93.054430,doi:10.1063/1.4789962} 
For a counter clockwise (CCW) spin wave this localization is on the opposite surface. If the magnetic properties are asymmetric along the film thickness, this results in different resonance frequencies for the CW and CCW spin wave, which leads to a frequency non-reciprocity as CW and CCW waves travel in opposite directions. In this letter, this asymmetry results from asymmetries in the magnetic anisotropy across the bulk of the film.

A typical device used to measure these spin waves is shown in~\figref{fig:figure1}b. Here two spin-wave antennas are placed on top of a magnetic strip. We drive an RF current through these antennas (whose spatial periodicity determines the spin-wave wavevector $k$), which excites spin waves through its time-dependent Oersted fields. These spin waves then propagate to the second antenna, where they are detected via induction ($L_\mathrm{xy}$). By inverting the detection and excitation antenna, we reverse the propagation direction of the detected spin waves. The magnetic strips consist of Ta(4)/[Pt/Ir](4)/Co(t)/[Pt/Ir](3)/Pt(2) and we vary the $k$ vector from $4$ to $10$~\si{\mu m^{-1}} in 1.5~\si{\mu m^{-1}} increments by varying the antenna geometry. The exact fabrication and measurement procedure is described in Ref.~\onlinecite{2019arXiv190111108L}.

\begin{figure*}
\centering
\includegraphics{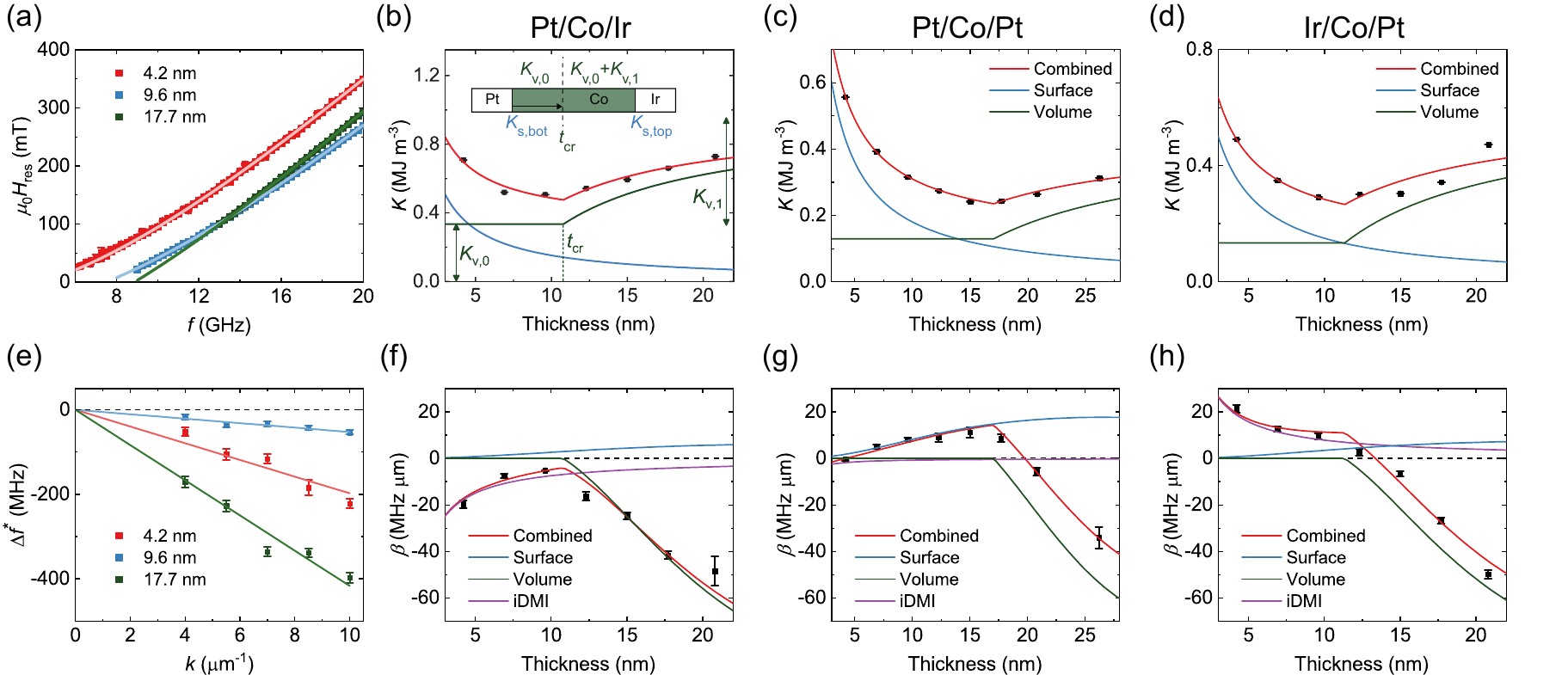}
\caption{\label{fig:figure2}(a) Fitted resonance fields $H_\mathrm{res}$ as a function of frequency $f$ for Pt/Co($t$)/Ir at different Co thicknesses $t$. (b-d) Anisotropy $K$ as a function of Co thickness $t$ for the three different stacks together with a fit that includes both a bulk and interfacial term which are plotted separately. The different parameters that determine the bulk contribution ($K_\mathrm{v,0}$, $K_\mathrm{v,1}$ and $t_\mathrm{cr}$) are labeled in (b). The inset of (b) shows a sideview of the magnetic stack with the different anisotropy components labeled. (e) Converted frequency shifts $\Delta f^{*}$ as a function of wavevector $k$ for Pt/Co($t$)/Ir at different Co thicknesses $t$, including a linear fit through the origin. (f-h) Slope $\beta$ of the wavevector dependence of the shift extracted from linear fits [see (e)] as a function of layer thickness for the three different stacks. Also included is a fit that models this shift (combined) and the individual components (iDMI, surface and bulk) of that fit. For the fit parameters of (b-d) see~\tabref{tab:table1} and for (f-h) see~\tabref{tab:table2}.}
\end{figure*}

We first investigate the self-induction $L_\mathrm{xx}$ of the antennas to extract the magnetic anisotropy. A typical measurement is shown in~\figref{fig:figure1}c, where $L_\mathrm{11}$ is plotted as a function of the magnetic field $H$. This spectrum shows a typical FMR-like resonance profile indicative of spin-wave excitation. The real and imaginary part are fitted simultaneously with a linear combination of a symmetric and anti-symmetric Lorentzian line-shape such that the resonance field $H_\mathrm{res}$ can be extracted (dashed line). 
Extracting the resonance fields for different frequencies and different Co thicknesses $t$ produces~\figref{fig:figure2}a. Here, the resonance fields are fitted using well-known Kittel-like relations, with only the out-of-plane (OOP) anisotropy $K$ as a fit parameter.~\cite{0022-3719-19-35-014,Kalinikos1981}\footnote{With $M_\mathrm{s}$=$1.44$~\si{MA.m^{-1}}, $g=2.17$, $k=7$~\si{\mu m^{-1}} (dictated by the antenna), and $w_\mathrm{eff}=1.2$~\si{\mu m}~\cite{2019arXiv190111108L}.}

In~\figref{fig:figure2}b-d we plot the fitted $K$ as a function of $t$ for the three different stacks. For all stacks, $K$ decreases for increasing $t$ when $t \lesssim 10$~\si{nm}. This is the interfacial anisotropy that reduces in magnitude due to the increasing magnetic volume. Above this thickness, we find that the anisotropy starts to increase again. This is attributed to a crystalline phase transition of the Co from face-centred cubic (fcc) to hexagonal close-packed (hcp) above a critical thickness $t_
\mathrm{cr}$, already widely observed in literature.~\cite{RIEDI199997,Johnson_1996,PhysRevLett.115.056601,PhysRevLett.81.5229,WELLER19951563,PhysRevB.99.184439} In the supplementary information we confirm the presence of different structural phase contributions in films with different thicknesses using transmission electron microscopy (TEM) images in conjunction with nuclear magnetic resonance (NMR) measurements. As the hcp phase has a much larger magneto-crystalline anisotropy along the c-axis (aligned along the OOP direction) this leads to an increase in $K$ along the OOP direction.~\cite{Johnson_1996} Both OOP anisotropy contributions can be fitted simultaneously as 
\begin{equation}
\label{eq:eq1}
K = \begin{cases}
\frac{K_\mathrm{s}}{t}+K_\mathrm{v,0} &\text{$t \leq t_{\mathrm{cr}}$}\\
\frac{K_\mathrm{s}}{t}+K_\mathrm{v,0}+K_\mathrm{v,1} \frac{t-t_\mathrm{cr}}{t} &\text{$t > t_{\mathrm{cr}}$}
\end{cases}
\end{equation}
with $K_\mathrm{s}=K_\mathrm{s,bot}+K_\mathrm{s,top}$ the total interfacial anisotropy, $K_\mathrm{v,0}$ the crystalline anisotropy of the bottom half of the Co film, and $K_\mathrm{v,1}$ the difference in the anisotropy between the top and bottom half of the film. This additional crystalline anisotropy is now included as a volume weighted average through the last term, where we assume an fcc phase of thickness $t_\mathrm{cr}$ in the bottom half of the film, with the remainder of the Co film in the hcp phase (see inset \figref{fig:figure2}b). In~\figref{fig:figure2}b we fit the data to~\verref{eq:eq1} and label the individual fitting parameters. The fits for the other two stacks are similarly plotted in~\figref{fig:figure2}c-d. 

\begin{table*}
\caption{\label{tab:table1}%
Fit parameters of the fits of the anisotropy for the different stacks displayed in~\figref{fig:figure2}b-d. They include the surface anisotropy $K_\mathrm{s}$ and the 3 volume anisotropy terms indicated in~\verref{eq:eq1}.}
\begin{ruledtabular}
\begin{tabular}{ldddd}
&
\multicolumn{1}{c}{$K_\mathrm{s}$~(\si{mJ.m^{-2}})}&
\multicolumn{1}{c}{$K_\mathrm{v,0}$~(\si{MJ.m^{-3}})}&
\multicolumn{1}{c}{$K_\mathrm{v,1}$~(\si{MJ.m^{-3}})}&
\multicolumn{1}{c}{$t_\mathrm{cr}$~(\si{nm})}\\
\colrule
Pt/Co/Ir & 1.5 \pm 0.3 &  0.33 \pm 0.05 &  0.63 \pm 0.07 & 10.8 \pm 0.6  \\
Pt/Co/Pt & 1.80 \pm 0.02 &  0.130 \pm 0.004 &  0.31 \pm 0.06 & 17.0 \pm 0.5 \\
Ir/Co/Pt & 1.5 \pm 0.2 &  0.13 \pm 0.04 &  0.5 \pm 0.1 & 11.3 \pm 0.9 \\
\end{tabular}
\end{ruledtabular}
\end{table*}
\begin{table*}
\caption{\label{tab:table2}%
Fit parameters from the fits of the slopes of the shifts for the different stacks shown in~\figref{fig:figure2}f-h. They include the terms that induce a shift, which is the increase in volume anisotropy $K_\mathrm{v,1}$ above $t_\mathrm{cr}$, the iDMI $D_\mathrm{s}$, and difference in surface anisotropies $\Delta K_\mathrm{s}=K_\mathrm{s,bot}-K_\mathrm{s,top}$. The last two columns use the $K_\mathrm{s}$ from~\tabref{tab:table1} and combines it with $\Delta K_\mathrm{s}$ to calculate the interfacial anisotropies at the bottom and top interface.}
\begin{ruledtabular}
\begin{tabular}{lddddd}
&
\multicolumn{1}{c}{$K_\mathrm{v,1}$~(\si{MJ.m^{-3}})}&
\multicolumn{1}{c}{$D_\mathrm{s}$~(\si{pJ.m^{-1}})}&
\multicolumn{1}{c}{$\Delta K_\mathrm{s}$~(\si{mJ.m^{-2}})}&
\multicolumn{1}{c}{$K_\mathrm{s,bot}$~(\si{mJ.m^{-2}})}&
\multicolumn{1}{c}{$K_\mathrm{s,top}$~(\si{mJ.m^{-2}})}\\
\colrule
Pt/Co/Ir &  0.30 \pm 0.03 & -1.0 \pm 0.2 & 0.2 \pm 0.1 & 0.9 \pm 0.2 & 0.7 \pm 0.2 \\
Pt/Co/Pt &  0.25 \pm 0.04 & -0.10 \pm 0.04 & 0.66 \pm 0.06 & 1.23 \pm 0.03 & 0.57 \pm 0.03\\
Ir/Co/Pt &  0.32 \pm 0.04 & 1.0 \pm 0.2 & 0.3 \pm 0.2 & 0.9 \pm 0.1 & 0.6 \pm 0.2\\
\end{tabular}
\end{ruledtabular}
\end{table*}
The resulting parameters from these fits are given in~\tabref{tab:table1}. For the crystalline volume anisotropy terms, we find that there is quite some variation between the different stacks. The variation in $t_\mathrm{cr}$ and $K_\mathrm{v}$ for Pt/Co/Ir and Pt/Co/Pt is hard to explain since both are grown on nominally identical underlayers, which should govern the behaviour of these parameters. We tentatively attribute this to different growth conditions, as Pt/Co/Ir was grown in a different batch from Pt/Co/Pt and Ir/Co/Pt. Yet, the values for $K_\mathrm{v}$ are in line with literature, where $K_\mathrm{v,1} \approx 0.5$~\si{MJ.m^{-3}}.~\cite{Johnson_1996} The values we find for $t_\mathrm{cr}$ are at least a factor $2\operatorname{-}3$ larger than those reported in literature for Pt/Co and Cu/Co systems. However, these details depend sensitively on the exact fabrication conditions.~\cite{RIEDI199997,PhysRevLett.115.056601,PhysRevLett.81.5229,WELLER19951563} 

With the anisotropy determined, we now focus our attention on the spin-wave transmission measurements to determine the frequency non-reciprocity. A typical transmission measurement of $L_\mathrm{xy}$ as function of magnetic field $H$ is shown in~\figref{fig:figure1}d. It shows a shift in resonance fields (dashed lines) $\Delta H$ between the oppositely propagating spin waves ($L_\mathrm{21}$ vs. $L_\mathrm{12}$) of about $2.4$~\si{mT}.\footnote{The shifts are determined using individual cross-correlations of the real and imaginary part of $L_\mathrm{12}$ with $L_\mathrm{21}$. They are then averaged with the negative shifts at negative fields to remove any biases. A method where we fit the actual peak locations was also used and yielded similar shifts (see the supplementary material). Some additional considerations on shift-extraction are also presented in the supplementary.} This field shift is converted to a frequency shift $\Delta f^{*}$ that is linear in $k$ and (mostly) independent of the applied magnetic field when looking at shifts due to iDMI and $\Delta K_\mathrm{s}$.~\cite{PhysRevB.88.184404,PhysRevB.93.054430} Similar to how ferromagnetic resonance linewidths are converted,~\cite{doi:10.1063/1.2197087} we calculate $\Delta f^{*}=-\left(\frac{\partial H_\mathrm{res}}{\partial f}\right)^{-1} \Delta H$. These shifts are plotted as a function of $k$ for arbitrary thicknesses in~\figref{fig:figure2}e. For all measurements the shifts are linear in $k$ and the fitted slope $\beta$ is used as a measure for the spin-wave frequency non-reciprocity.

As a final step in the analysis, in~\figref{fig:figure2}f we plot $\beta$ as a function of layer thickness for Pt/Co/Ir. $\beta$ is negative for all thicknesses and decreases as $\sim 1/t$ up to $t \approx 10$~\si{nm}, in agreement with an iDMI contribution that decreases with increasing thickness. We attribute the increase in $\beta$ at $t=t_\mathrm{cr}$ to the increase in crystalline anisotropy for $t>t_\mathrm{cr}$. As the spin waves are localized at one of the two interfaces, the fact that the top part of the Co has a different crystalline volume anisotropy should indeed lead to a non-reciprocity, very similar to a non-reciprocity induced by a difference in surface anisotropies. In the supplementary material we derive an analytical equation that we fit to $\beta$ in~\figref{fig:figure2}f. This fit contains $3$ contributions; i. the iDMI which decreases as $1/t$. ii. a surface contribution due to $\Delta K_\mathrm{s}=K_\mathrm{s,bot}-K_\mathrm{s,top}$ which increases as $t^2$~\cite{PhysRevB.93.054430}, and iii. the bulk volume contribution stemming from a different crystalline anisotropy above $t_\mathrm{cr}$. Using the results from the fit of~\figref{fig:figure2}b, the shifts were fitted with $K_\mathrm{v,1}$, $D_\mathrm{s}$ and $\Delta K_\mathrm{s}$ as the free parameters. As demonstrated in~\figref{fig:figure2}f, there is an excellent agreement between the model and the measured shifts. Moreover, we find that the shift is dominated by the iDMI below $t_\mathrm{cr}$ and by the volume term due to the crystal phase transition above $t_\mathrm{cr}$. This is in contrast to literature, where a non-reciprocity at higher thicknesses is usually ascribed to differences in surface anisotropies.~\cite{Lee2016,PhysRevB.93.054430} The slopes $\beta$ and corresponding fits for Pt/Co/Pt and Ir/Co/Pt are shown in~\figref{fig:figure2}g and h and the resulting fit parameters of the shifts are displayed in~\tabref{tab:table2}.

With these results, we make three observations. First, there is the expected behaviour of the effective iDMI, which changes sign upon stack reversal between Pt/Co/Ir and Ir/Co/Pt. Moreover, for the nominally symmetric Pt/Co/Pt stack the iDMI is heavily reduced, as expected because the global inversion symmetry is no longer broken.~\cite{PhysRevB.78.140403,PhysRevLett.87.037203} From literature, the sign of the iDMI at the Pt/Co interface is well known, but there is still intense debate about the sign of the iDMI at the Ir/Co interface.~\cite{2019arXiv190205523K} Because the iDMI in Pt/Co/Ir stack is enhanced with respect to Pt/Co/Pt, we know the iDMI at the Ir/Co interface is either much smaller and/or has the opposite sign with respect to a Pt/Co interface. Additionally, the negligible DMI of the Pt/Co/Pt stack indicates that the DMI at the Pt/Co and Co/Pt interface is almost equal. Combining this with an iDMI for Pt/Co/Ir and Ir/Co/Pt that is smaller than the expected DMI at the Pt/Co interface of $\approx -1.5$~\si{pJ.m^{-1}}~\cite{doi:10.1021/acs.nanolett.6b01593} suggests that in our system the sign at the Ir/Co interface is the same as that of the Pt/Co interface.~\cite{2019arXiv190205523K}

Second, the differences in surface anisotropies are of the same sign such that the bottom interface always has a higher anisotropy than the top interface. The last two columns in~\tabref{tab:table2} calculate the corresponding interfacial terms, where we find that the Pt/Co and Ir/Co interface have approximately the same interfacial anisotropy, but that the bottom surface always has a higher anisotropy compared to the corresponding top interface, confirming earlier conjectures.~\cite{PhysRevB.91.104414,6094288} If we assume that both the anisotropy and iDMI depend in a similar matter on the interfacial quality, we can extrapolate the ratio between $K_\mathrm{s,bot/top}$ to the iDMI for Pt/Co/Pt. This gives an iDMI at the bottom Pt/Co interface of about $-0.2$~\si{pJ.m^{-1}}. As this is significantly lower than what is reported ($-1.5$~\si{pJ.m^{-1}}~\cite{doi:10.1021/acs.nanolett.6b01593}), it would suggest that the iDMI and anisotropy do not depend in a similar matter on the interfacial quality.

Last, the values for $K_\mathrm{v,1}$ (\tabref{tab:table2}) can vary by a factor of $2$ from the results of the anisotropy fits (\tabref{tab:table1}). The TEM and NMR data show a gradual transition between the fcc and hcp phase as a function of thickness. In contrast, the assumed anisotropy profile (eq.~\ref{eq:eq1}) describes an instantaneous transition from fcc to hcp at $t_\mathrm{cr}$. This oversimplification in the fits could potentially explain the different $K_\mathrm{v,1}$ values.

We have shown that PSWS can be used to extract the different contributions to the frequency non-reciprocity over a wide thickness range. This makes it an extremely powerful tool for fundamental investigations into the DMI. For example, although very little experimental work has been done in this direction, there is great interest in the manipulation of the iDMI via an electric field (EF).~\cite{Srivastava2018,*doi:10.1063/1.5050447,*Nawaoka_2015} PSWS should prove very powerful in quantifying the effect of the EF on the DMI,~\cite{Lee2016} as it is able to separate the EF effect on the iDMI from the EF effect on the anisotropy. The latter is known to be present and, as we demonstrate, cannot be ignored when interpreting the frequency non-reciprocity to extract the iDMI.~\cite{*[{See for example }] [{ and references therein.}] Rana2019}
The additional effects demonstrated here could also explain some of the puzzling behaviour in Ref.~\onlinecite{Lee2016}, where PSWS was used to measure iDMI in thick films of Pt/Co/MgO. Here, the iDMI-induced shift seems to be of the wrong sign and significantly larger than reported elsewhere in literature.~\cite{Lee2016,Nembach2015,Cho2015,PhysRevLett.114.047201,doi:10.1021/acs.nanolett.6b01593} 

The large non-reciprocity demonstrated in this letter, induced by the crystalline phase change, can also be used in the field of magnonics. Different types of (proposed) devices rely extensively on spin-wave non-reciprocity of some kind.~\cite{CAMLEY1987103,PhysRevLett.117.197204,doi:10.1063/1.4819435,Jamali2013} Although iDMI can enhance this non-reciprocity~\cite{Lee2016,Nembach2015,Cho2015,PhysRevLett.114.047201,doi:10.1021/acs.nanolett.6b01593} the thin films required to generate large non-reciprocities usually have large damping and low spin-wave group velocities. Rather, this work suggests that using crystalline anisotropies might offer a significantly more practical route towards increasing the spin-wave non-reciprocity. 
Although the system investigated here relies on a strain-induced crystalline phase transition that can be impractical, more feasible routes can be imagined; for instance, using a bilayer of fcc Co and [Co/Ni] repeats~\cite{PhysRevB.96.024401} to act as the low and high anisotropy materials respectively. This additionally leads to a naturally occurring magnetization gradient across the thickness, further enhancing the frequency non-reciprocity.~\cite{Gallardo_2019}


Summarizing, we have shown in this letter that the physics behind spin-wave frequency non-reciprocity is more complex than originally assumed and includes a yet unnoticed but important contribution that is the result of a change in structural phase as function of film thickness. However, by investigating the thickness dependence of the non-reciprocity we can uniquely isolate the iDMI, the difference in interfacial anisotropies and a large contribution induced by this crystalline phase transition. 

\begin{acknowledgments}
This work is part of the research programme of the Foundation for Fundamental Research on Matter (FOM), which is part of the Netherlands Organisation for Scientific Research (NWO). Solliance and the Dutch province of Noord-Brabant are acknowledged for funding the TEM facility. Financial support is acknowledged from the Deutsche Forschungsgemeinschaft (DFG) through Grants No. WU595/3-3, and WU595/14-1. We also thank V. Vandalon for help with the quantitative TEM analysis.
\end{acknowledgments}
\bibliography{references}
\end{document}


\title{Supplementary Material: Extraction of Dzyaloshinksii-Moriya interaction from propagating spin waves validated}
\author{Juriaan Lucassen}
\email{j.lucassen@tue.nl}
\affiliation{Department of Applied Physics, Eindhoven University of Technology, P.O. Box 513, 5600 MB Eindhoven, the Netherlands}
\author{Casper F. Schippers}
\affiliation{Department of Applied Physics, Eindhoven University of Technology, P.O. Box 513, 5600 MB Eindhoven, the Netherlands}
\author{Marcel A. Verheijen}
\affiliation{Department of Applied Physics, Eindhoven University of Technology, P.O. Box 513, 5600 MB Eindhoven, the Netherlands}
\affiliation{Eurofins Materials Science BV, High Tech Campus 11, 5656 AE Eindhoven, The Netherlands}
\author{Patrizia Fritsch}
\affiliation{IFW-Dresden, Institute for Solid State Research, Helmholtzstraße 20, 01069 Dresden, Germany}
\author{Erik Jan Geluk}
\author{Beatriz Barcones}
\affiliation{NanoLab@TU/e, Eindhoven University of Technology, P.O. Box 513, 5600 MB Eindhoven, the Netherlands}
\author{Rembert A. Duine}
\affiliation{Department of Applied Physics, Eindhoven University of Technology, P.O. Box 513, 5600 MB Eindhoven, the Netherlands}
\affiliation{Institute for Theoretical Physics, Utrecht University, Princetonplein 5, 3584 CC Utrecht, the Netherlands}
\author{Sabine Wurmehl}
\affiliation{IFW-Dresden, Institute for Solid State Research, Helmholtzstraße 20, 01069 Dresden, Germany}
\affiliation{Institute of Solid State and Materials Physics, TU Dresden, 01062 Dresden, Germany}
\author{Henk J.M. Swagten}
\affiliation{Department of Applied Physics, Eindhoven University of Technology, P.O. Box 513, 5600 MB Eindhoven, the Netherlands}
\author{Bert Koopmans}
\affiliation{Department of Applied Physics, Eindhoven University of Technology, P.O. Box 513, 5600 MB Eindhoven, the Netherlands}
\author{Reinoud Lavrijsen}
\affiliation{Department of Applied Physics, Eindhoven University of Technology, P.O. Box 513, 5600 MB Eindhoven, the Netherlands}

\date{\today}

\maketitle
\section{Frequency dependence of shifts}
In this section we demonstrate that one should be careful when extracting the peak-shifts from the transmission data for two reasons (which will be addressed in the following paragraphs):
\begin{enumerate}
	\item Due to the dipolar fields within the magnetic strip, the magnetization in the strip can be inhomogeneous if the magnetic field $H$ is not large enough to saturate the magnetization. 
	\item There can be a parasitic coupling between the two antennas that can result in an extracted peak shift of the wrong sign.
\end{enumerate}

\begin{figure*}
\centering
\includegraphics[width=1\textwidth]{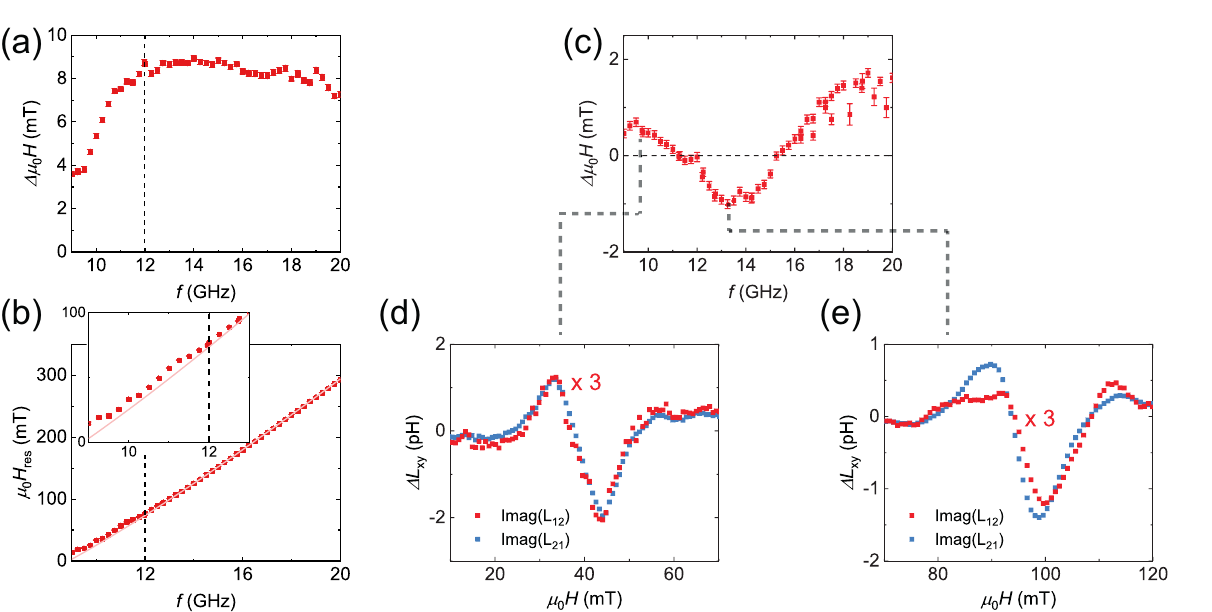}
\caption{\label{fig:freq_dep} (a-b) Data for Pt/Co(17.7)/Ir measured with $k=7$~\si{\mu m^{-1}} antennas where the dashed line indicates the frequency below which we the behaviour starts deviating from expectations. (a) Extracted field shifts a function of frequency. (b) Resonance field $H_\mathrm{res}$ as a function of $f$. Inset shows a blown up version at low frequencies. (c-e) Data for Pt/Co(9.6)/Ir measured with  $k=7$~\si{\mu m^{-1}} antennas. Shifts a function of frequency (c) and mutual-induction $\Delta L_\mathrm{xy}$ as a function of applied magnetic field $H$ measured at $9.78$~\si{GHz} (d) and $12.75$~\si{GHz} (e) where we rescaled $L_\mathrm{12}$ to demonstrate the relative behaviour.}
\end{figure*}

The problem of inhomogeneous magnetization shows up when we look at the dispersion relation shown in~\figref{fig:freq_dep}b. Here, we find that below $\approx 12$~\si{GHz} the dispersion relation no longer describes the measured data properly. We believe that this is a result of an inhomogeneous magnetization profile due to the internal dipolar fields, as described in Refs.~\onlinecite{doi:10.1063/1.101131,PhysRevLett.91.137204}. At these same frequencies, the extracted peak shifts are shown in~\figref{fig:freq_dep}a, where we find a constant shift except for $f<12$~\si{GHz}, when the dispersion relation starts deviating. This means that when extracting peak-shifts, we must ensure that the magnetic strip is saturated by working at high enough fields (and thus frequencies), determined by looking at when the dispersion relationship fits properly to the measured resonance fields. The saturation field scales with the thickness of the film~\cite{{doi:10.1063/1.101131,PhysRevLett.91.137204}} which is also what we find; as the film thickness decreases we can work at smaller fields/frequencies.

The second problem has to do with a parasitic coupling, described in full detail in Ref.~\onlinecite{2019arXiv190111108L}. Because the antenna-antenna spacing and the amount of meanders is reduced, the issue is not as significant here, but can be present nonetheless. For example, in~\figref{fig:freq_dep}c the shifts as a function of frequency for Pt/Co(9.6)/Ir are shown. We find a strong frequency dependent shift that is not related to an inhomogeneous magnetization profile described earlier. Instead, it is now related to this parasitic coupling. In~\figref{fig:freq_dep}d~and~\figref{fig:freq_dep}e the mutual inductance spectra from which the shifts were extracted are shown. We find that at low frequencies (d) the spectra have exactly the same shape, but are shifted just slightly with respect to each other. However, at larger frequencies (e) this is no longer the case and the $L_\mathrm{12}$ spectrum is contaminated by a parasitic coupling. This complicates the shift extraction to the point that we get an artificial sign change. Fortunately, this behaviour is only an issue when the intrinsic shifts are very small, which is only at a few selected thicknesses (see Fig.~2f-h of the main paper). Furthermore, it requires that the spin wave transmission signal be small, which is at low thicknesses where the group velocity and the attenuation length are reduced, and when the frequency is relatively high (in the example above $10$~\si{GHz}). However, by measuring at low frequencies this issue can be resolved.

In both cases, we have also used the model as described in~\secref{sec:peak_shift} to find out if this frequency dependence did not result from any field (or frequency) dependence in the shifts. Using the parameters from Pt/Co/Ir given in the main paper, we find that there is some frequency/field dependence of the non-reciprocity; approximately $90$~\si{MHz} going from $0$ to $300$~\si{mT} at $20$~\si{nm} (about $20$~\%), and $0.2$~\si{MHz} at $10$~\si{nm} (about $5$~\%). However, these variations are nowhere large enough to explain the anomalous behaviour seen in this section, where we find variations above $100$~\%.
\section{Shifts by determining the peak position}
In this section we show the slope data extracted from the shifts determined by determining the actual spin wave peak locations in the mutual inductance. We show that this method of shift extraction produces results similar to the cross-correlation method used in the main paper, confirming the correctness of the cross-correlation method. 

\begin{figure*}
\centering
\includegraphics{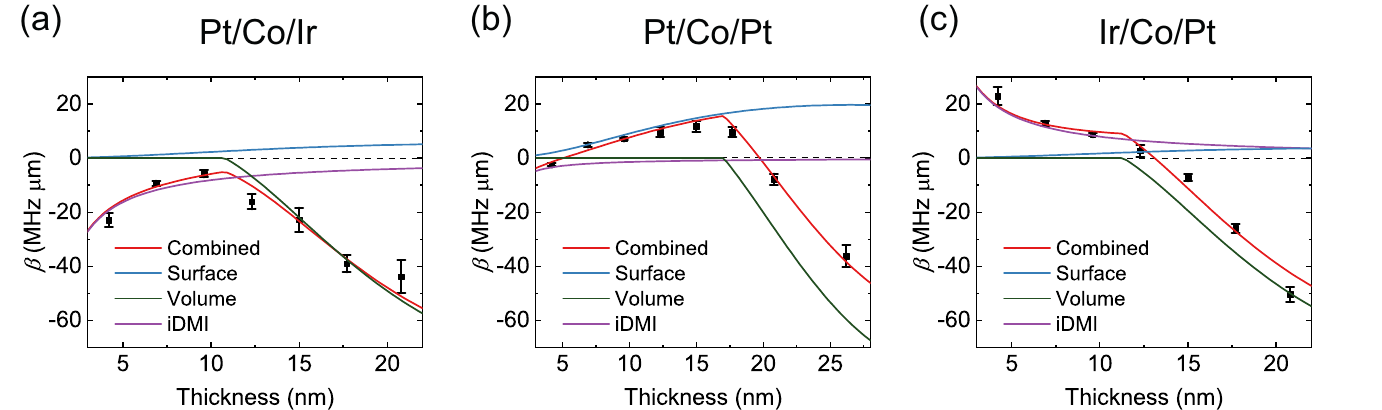}
\caption{\label{fig:peak_pos} (a-c) Slope of the wavevector dependence of the shift extracted from linear fits as a function of layer thickness for the three different stacks. Also included is a fit that models this shift (combined) and the individual components (surface, volume an iDMI) of that fit. Compared to Fig.~2 of the main paper, here the peak shifts from the mutual inductance data were determined not by cross-correlation, but by actually determining the peak positions. For the fit parameters see~\tabref{tab:table2}.}
\end{figure*}

Similar to Fig.~2f-h and Table~2 of the main paper, we show in~\figref{fig:peak_pos} the slopes as a function of layer thickness and in~\tabref{tab:table2} the resulting fit-parameters. We find that within error bars the results of~\tabref{tab:table2} match the results of Table~2 of the main paper, demonstrating that both methods can be used to reliably extract peak shifts. 

\begin{table*}[!h]
\caption{\label{tab:table2}%
Fit parameters from the fits of the slopes of the shifts for the different stacks shown in~\figref{fig:peak_pos}a-c. They include the terms that induce a shift, which is the increase in volume anisotropy $K_\mathrm{v,1}$ above $t_\mathrm{cr}$, the iDMI $D_\mathrm{s}$, and difference in surface anisotropies $\Delta K_\mathrm{s}$. The last two columns use the $K_\mathrm{s}$ from Table~1 of the main paper and combines it with $\Delta K_\mathrm{s}$ to calculate the interfacial anisotropies at the bottom and top interface. }
\footnotesize
\begin{ruledtabular}
\begin{tabular}{lddddd}
&
\multicolumn{1}{c}{$K_\mathrm{v,1}$~(\si{MJ.m^{-3}})}&
\multicolumn{1}{c}{$D_\mathrm{s}$~(\si{pJ.m^{-1}})}&
\multicolumn{1}{c}{$\Delta K_\mathrm{s}$~(\si{mJ.m^{-2}})}&
\multicolumn{1}{c}{$K_\mathrm{s,bot}$~(\si{mJ.m^{-2}})}&
\multicolumn{1}{c}{$K_\mathrm{s,top}$~(\si{mJ.m^{-2}})}\\
\colrule
Pt/Co/Ir &  0.26 \pm 0.04 & -1.0 \pm 0.2 & 0.2 \pm 0.2 & 0.9 \pm 0.2 & 0.7 \pm 0.2 \\
Pt/Co/Pt &  0.27 \pm 0.03 & -0.19 \pm 0.06 & 0.73 \pm 0.07 & 1.27 \pm 0.04 & 0.54 \pm 0.04\\
Ir/Co/Pt &  0.29 \pm 0.04 & 1.0 \pm 0.3 & 0.1 \pm 0.3 & 0.8 \pm 0.2 & 0.7 \pm 0.2\\
\end{tabular}
\end{ruledtabular}
\end{table*}
\section{NMR data}
As outlined in the main part of this work, different anisotropy contributions were needed to describe and fit the data from Propagating Spin Wave Spectroscopy (PSWS) and to extract the interfacial Dzyaloshinskii–Moriya interaction (iDMI), specifically relying on a proposed crystalline phase transition between fcc and hcp Co as a function of thickness. In this section, we describe how zero-field nuclear magnetic resonance ($^{59}$Co NMR) measurements can determine the relative amounts of different crystallographic Co phases. Specifically, two Ta(4)/Pt(4)/Co(t)/Ir(3)/Pt(2) films with $t=10,25$~\si{nm} were investigated, where one stack has a thickness below $t_\mathrm{cr}$ and one above (for details see main part of this work).

Zero-field NMR is well known as excellent tool to locally probe different crystallographic environments within Co thin films.~\cite{pan97,RIEDI199997,Strijkers2000,WJK05a,WJK05b} The local sensitivity of zero-field NMR arises from the local contributions to the hyperfine field which in turn strongly depend on the local crystallographic environments of the NMR active nuclei, hence on their surrounding neighbouring atoms, as well as their site symmetry and magnetic properties, their degree of order and other crystallographic defects. For more details the interested reader is referred to several review articles and references therein.~\cite{pan97,WK08}
The $^{59}$Co NMR measurements were performed at a temperature of $5$~\si{K} in an automated, coherent, phase-sensitive, and frequency-tuned spin-echo spectrometer (NMR Service, Erfurt, Germany). In order to increase signal intensity, which scales with the number of probed $^{59}$Co nuclei, the films were cut in two parts; the resulting two slices were sandwiched with the substrate sides facing each other. All resonance spectra were recorded over a frequency range of $190$-$240$~\si{MHz} with a step size of $0.5$~\si{MHz}, with a $90^{\circ}$ - $90^{\circ}$ pulse of $0.6$~\si{\mu s} pulse length and a $T_{1}$ of $100$~\si{ms}. The recorded $^{59}$Co NMR spectra were corrected for the enhancement factor (see, e.g. Refs.~\onlinecite{pan97,RIEDI199997,WK08}) and for the $\nu^{2}$-dependence. 

\begin{figure}[!h]
 \centering
 	\includegraphics[width=0.6\columnwidth]{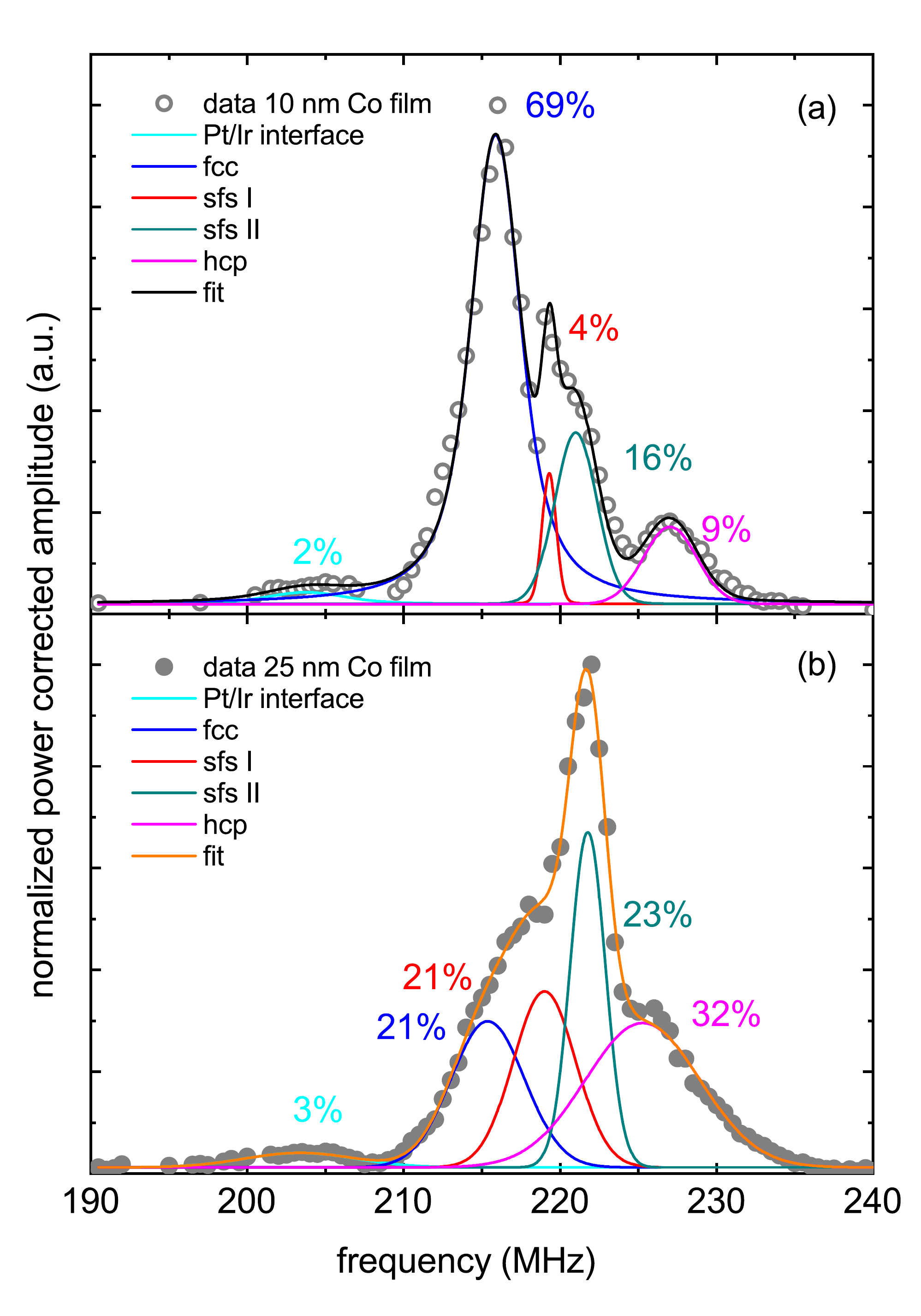}
	\label{fig:spec}
 	\caption{$^{59}$Co NMR spectra for (a) the film with Co thickness $10$~\si{nm} and (b) $25$~\si{nm}. Both spectra were fitted with a total of five Gaussian lineshapes, representing the different environments of Co nuclei: the Co atoms at the interface to the Pt and Ir layers, Co in an fcc environment, Co in a hcp environment, and two distinct stacking faults sfs I and II.  The percentages correspond to the relative area of each Gaussian line to the overall spectrum.
 	}
 \end{figure}

In~\figref{fig:spec}a and b we plot the NMR frequency spectra for the $10$~and~$25$~\si{nm} samples, respectively. Each spectrum consists of $5$ resonances originating from different different local environments of the Co nuclei, and each is fitted with a Gaussian line shape (also shown).
The area of each Gaussian line corresponds directly to the number of nuclei in the respective local environment, thus quantifying the corresponding contributions of specific local Co environments (percentages in figure).~\cite{Strijkers2000,WK08,pan97} To understand the different contributions, we discuss them individually:

\begin{itemize}
	\item The resonance line with the lowest frequency has only a small contribution to the overall spectrum and stems from Co nuclei in the vicinity to the Pt and Ir layers of the film stack (cyan lines in \figref{fig:spec}). Since we observe a single distinct line, these interfaces are well and smoothly ordered.~\cite{Strijkers2000}

\item The resonance line at the highest frequency roughly corresponds to the frequency reported in literature for hexagonally closed packed (hcp) Co with the $c$-axis perpendicular to the magnetization direction (hcp$_{\bot}$).\cite{Strijkers2000} 

\item The line at $215$~\si{MHz} originates from Co in a face centered cubic (fcc) stacking structure, in line with previous reports on Co films.\cite{Strijkers2000} 

\item The two remaining lines arise from stacking faults (sfs). 
Unfortunately, the exact state of the measured stacking fault environment cannot be resolved by zero-field NMR alone.\cite{deGronckel1989} 
\end{itemize}

From the relative areas of the different contributions, we find that the hcp crystal structure is hardly present in the $10$~\si{nm} film, but at $25$~\si{nm} a third of the film is in a hcp environment. Similar to the hcp environment, we also find an increase in the number of stacking faults with increasing Co thickness. With the fcc environment we find the opposite; its relative presence decreases with increasing Co thickness. 

All this suggests that the $10$~\si{nm} film (below the threshold thickness $t_{\mathrm{cr}}$) is mainly of fcc type structure, while with increasing film thickness, the influence of the lattice mismatch from the substrate transforms the structure to hcp type stacking via formation of stacking faults. Hence, the NMR results strongly support the conclusion that the different magnetic anisotropy originates in the different crystallographic structure of films above and below a critical thickness $t_\mathrm{cr}$. 

\section{TEM data}
\begin{figure*}[!h]
\centering
\includegraphics[width=1.0\textwidth]{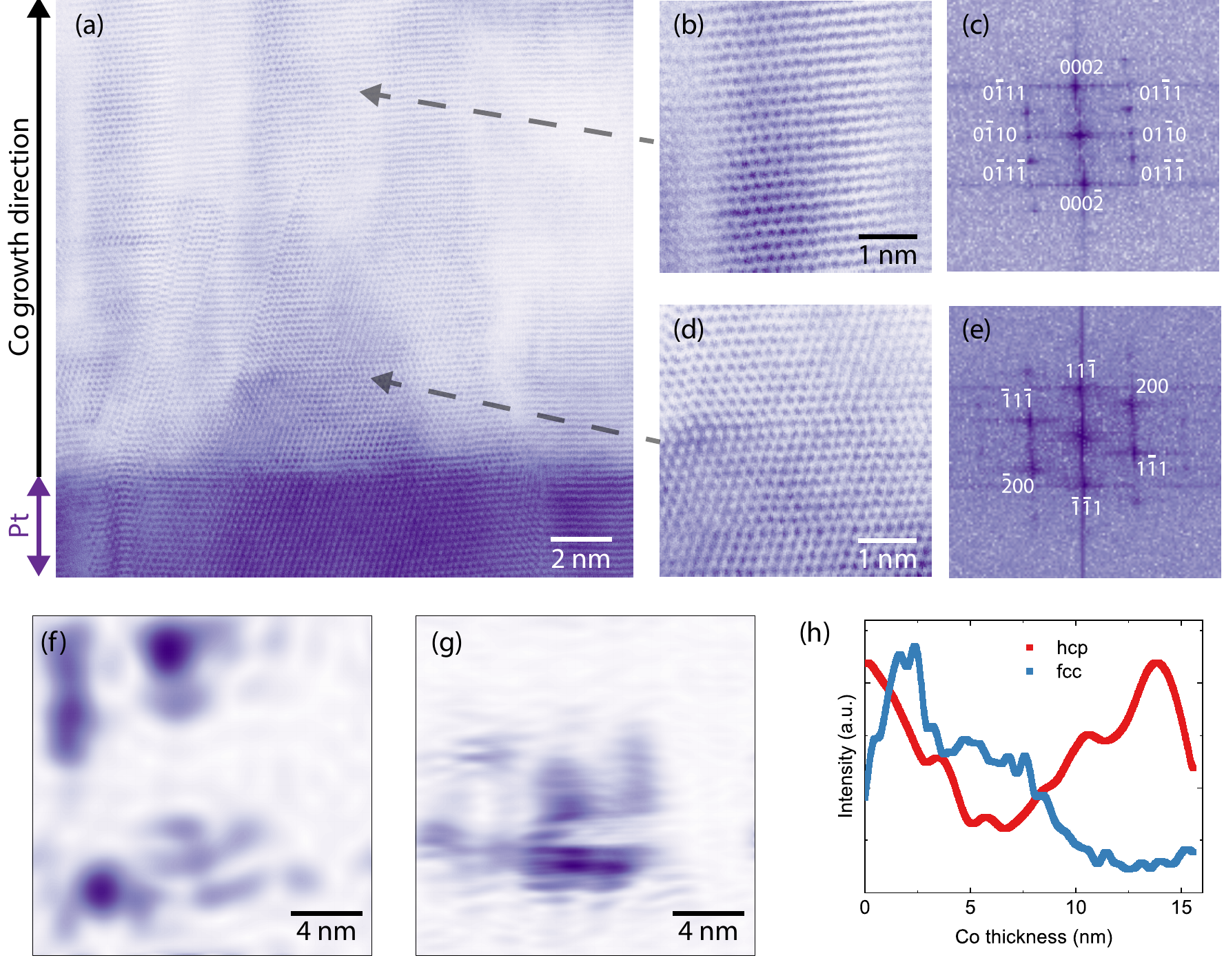}
\caption{\label{fig:TEM_1}(a) Cross section HAADF-STEM image (shown in inversed contrast) of Si/SiO2(100nm)//Ta(4)/Pt(4)/Co(30)/Ir(3)/Pt(2). (b,d) Two specific areas already shown in (a)  indicated by arrows but on an enlarged scale. The corresponding reciprocal images at these positions were calculated by FFT and are show in (c,e) demonstrating that a part of the film is hcp ordered with a ${<}0001{>}$ growth direction (c), and another part of the film part is fcc ordered with a ${<}11\bar{1}{>}$ growth direction (e).~\cite{Edington1975} (f-g) Masked and smoothed version of (a), where we show in (f) the hcp phase and in (g) the fcc phase. (h) Relative intensities of both the hcp and fcc phase as a function of Co layer thickness extracted from (f) and (g) by averaging out the lateral variations.}
\end{figure*}

In this section we confirm, using cross-section Transmission Electron Microscopy (TEM) that there is indeed an increase in the Co hcp phase fraction above $t_\mathrm{cr} \sim 10$~\si{nm}. Although strongly suggested by both the anisotropy data (see main paper) and the NMR data (see previous section), we can directly image this transition in $30$~\si{nm} thick Co films using TEM. We image a lamella of Si/SiO2(100nm)//Ta(4)/Pt(4)/Co(30)/Ir(3)/Pt(2) using a probe corrected JEOL ARM 200F Transmission Electron Microscope, equipped with a $100$~\si{mm^{2}} CenturioSDD EDX detector. It is operated at 200 kV with the images taken using a HAADF (High Angle Annular Dark Field) detector in Scanning TEM (STEM) mode.

A typical image is shown in~\figref{fig:TEM_1}a, which shows a Co layer on top of a polycrystalline Pt(111) seeding layer. Because of the polycrystalline nature, not all grains are aligned along the viewing direction. However, we can distinguish between two crystallographic phases in Co. In~\figref{fig:TEM_1}b we indicate the hcp phase, which grows in the ${<}0001{>}$ direction as indicated by the diffraction spots in~\figref{fig:TEM_1}c.~\cite{Edington1975} The second phase is an fcc phase growing in the ${<}111{>}$ direction as shown in~\figref{fig:TEM_1}d and its reciprocal image in~\figref{fig:TEM_1}e.~\cite{Edington1975} 

We mask the Fourier transform of~\figref{fig:TEM_1}a using the spots from~\figref{fig:TEM_1}c (except for the $01{-}11$ spots). 
After transforming this back, and smoothing the image the result is~\figref{fig:TEM_1}f which indicates which part of the Co is in the hcp phase. Doing something similar for the fcc phase, but now using the $(111)$ and $(200)$ spots (see~\figref{fig:TEM_1}e), results in~\figref{fig:TEM_1}g. Averaging along the lateral direction to get the intensities of the fcc and hcp phase as a function of Co layer thickness gives~\figref{fig:TEM_1}h. We find that the fraction of fcc phase does indeed decrease with increasing Co layer thickness. The hcp behaviour is more complex, where we find an initial decrease and then an increase above $5$~\si{nm}. The existence of hcp close to the interface can be understood from the presence of horizontal stacking faults in the cubic phase. These stacking faults are in fact single layer hexagonally stacked layers. The presence of this type of stacking fault is a common phenomenon in cubic close packed lattices.

\begin{figure*}
\centering
\includegraphics[width=0.7\textwidth]{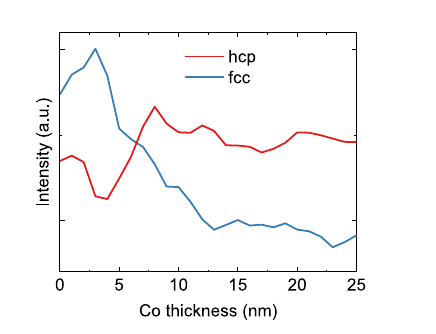}
\caption{\label{fig:TEM_2}Relative intensities of the hcp and fcc phases as a function of Co layer thickness averaged over several TEM images (similar to~\figref{fig:TEM_1}h).}
\end{figure*}
Because of the limited number of grains present in a single image, we perform the analysis shown in~\figref{fig:TEM_1} for multiple images and average the resulting profiles to get~\figref{fig:TEM_2}. Here, we find confirmation that indeed, there is a thickness $t_\mathrm{cr}$ of $\sim 5{-}10$~\si{nm} above which there is an increase in the hcp phase and a reduction of the fcc phase. Specifically, the hcp phase fraction increases between $5$ and $10$~\si{nm}; at about $10$~\si{nm}, the influence of the substrate is negligible, and, hence, the hcp phase stabilizes up from this thickness.~\cite{RIEDI199997,PhysRevLett.115.056601,PhysRevLett.81.5229,WELLER19951563} 

The TEM data allow us to extract this qualitative picture of the evolution of crystallographic phase fractions while a quantitative analysis is challenging. For one, the limited number of grains that are aligned along the viewing direction make it difficult to properly normalize an individual scan. Second, the close proximity of the hcp and fcc diffraction spots complicates the masking, leading to errors in the intensities. Thirdly, single layer stacking faults, as for example visible centrally in~\figref{fig:TEM_1}d, make the Fourier filtered images less binary. The two stacking faults in~\figref{fig:TEM_1}d can be considered as single-layer hexagonally stacked layers in an fcc crystal, and will appear as such in the filtered images. Lastly, we note that the critical thickness $t_\mathrm{cr}\sim 10$~\si{nm} agrees with the results of the main paper as well as the NMR data, but is a factor $\sim 2$ larger than those found in literature for Pt/Co and Cu/Co systems. However, these
details depend sensitively on the exact fabrication conditions.~\cite{RIEDI199997,PhysRevLett.115.056601,PhysRevLett.81.5229,WELLER19951563}
%
%
\section{Shift derivation}
\label{sec:peak_shift}
In this section, we will derive the resulting frequency non-reciprocity due to a difference in surface anisotropies as well as the following profile of out-of-plane (OOP, along the $z$ direction) crystalline anisotropy:
\begin{equation}
\label{eq:K}
K=K_\mathrm{v,0}+u(z-t_\mathrm{cr})K_\mathrm{v,1},
\end{equation}
with $z$ the position along the thickness $t$, $t_\mathrm{cr}$ the critical thickness above which Co transforms from fcc to hcp, $K_\mathrm{v,0}$ the magneto-crystalline anisotropy of the fcc phase, $K_\mathrm{v,1}$ the difference between the magneto-crystalline anisotropy of hcp and fcc phase, and $u(x)$ the step function.

The approach will be similar to that described in Refs.~\onlinecite{PhysRevB.93.054430,doi:10.1063/1.4789962}, where we solve the following linearised LLG equation for waves propagating in the $x$-direction (wavector $k$) with a magnetic field $H_\mathrm{0}$ applied in the $y$-direction
\begin{equation}
\label{eq:LLG}
i\omega \boldsymbol{m}= \gamma \mu_\mathrm{0} H_\mathrm{0} \boldsymbol{\hat{y}} \times \boldsymbol{m}- \gamma \mu_\mathrm{0} M_\mathrm{s} \boldsymbol{\hat{y}} \times \boldsymbol{h},
\end{equation}
where $\boldsymbol{m}=(m_\mathrm{x},0,m_\mathrm{z})$, $\boldsymbol{h}= (h_\mathrm{x},0,h_\mathrm{z})$ are the dynamic components of the magnetization and effective field respectively. $\gamma$ is the gyromagnetic ratio, $\mu_\mathrm{0}$ the permeability of vacuum and $M_\mathrm{s}$ the saturation magnetization. The total dynamic field can also be written, where we now explicitly introduce the $z$-dependence that occurs through localization, as
\begin{equation}
\begin{split}
\boldsymbol{h}(z)&=\frac{2A}{\mu_\mathrm{0} M_\mathrm{s}^2}\left(\frac{\partial^2}{\partial z^2}-k^2 \right) \boldsymbol{m}(z)+\int_0^t \mathrm{d}z' \bar{G}_k(z-z') \boldsymbol{m}(z')\\
&+\frac{2}{\mu_\mathrm{0} M_\mathrm{s}^2} \boldsymbol{\hat{z}} \bigl[ K_\mathrm{s,bot} \delta (z) m_\mathrm{z}(0)+ K_\mathrm{s,top} \delta (z-t) m_\mathrm{z}(t)+K_\mathrm{v,0} m_\mathrm{z}(z) +u(z-t_\mathrm{cr})K_\mathrm{v,1} m_\mathrm{z}(z) \bigr],
\end{split}
\end{equation}
with $A$ the exchange constant and $\bar{G}_k$ the magneto static Green's function.~\cite{doi:10.1063/1.4789962} The first term describes to the contribution to the effective field as a result from the exchange interaction, the second from the dipolar interactions and the third from the anisotropies, where we included the bulk term from~\verref{eq:K} and where $K_\mathrm{s,bot/top}$ is the interfacial anisotropy from the bottom/top interface ($K_\mathrm{s}=K_\mathrm{s,bot}+K_\mathrm{s,top}$ and $\Delta K_\mathrm{s}=K_\mathrm{s,bot}-K_\mathrm{s,top)}$). If we now expand $m$ onto the first two standing spin wave modes and insert everything into~\verref{eq:LLG} the result is the following eigenvalue equation
\begin{equation}
\label{eq:Mat}
i \Omega \begin{bmatrix}
m_\mathrm{x,0}\\
m_\mathrm{z,0} \\
m_\mathrm{x,1}\\
m_\mathrm{x,1}
\end{bmatrix}=\begin{bmatrix}
0 & \Omega_\mathrm{0,z} & -i Q & 0\\
-\Omega_\mathrm{0,x} & 0 & \delta & i Q \\
iQ & 0 & 0 & \Omega_\mathrm{1,z}\\
\delta & -i Q & -\Omega_\mathrm{1,x} & 0
\end{bmatrix}
\begin{bmatrix}
m_\mathrm{x,0}\\
m_\mathrm{z,0} \\
m_\mathrm{x,1}\\
m_\mathrm{x,1}
\end{bmatrix},
\end{equation}
where $m_\mathrm{i,j}$ is the expansion coefficient of the magnetization component in the i direction of the jth order standing spin wave mode, $\Omega=\frac{\omega }{\gamma \mu_\mathrm{0} M_\mathrm{s}}$ and
\begin{equation*}
\begin{split}
\Omega_\mathrm{0,x}&=1-P_\mathrm{00}-\epsilon_0-\epsilon_1+h+\Lambda^2k^2,\\
\Omega_\mathrm{0,z}&=h+P_\mathrm{00}+k^2 \Lambda^2,\\
\Omega_\mathrm{1,x}&=1-P_\mathrm{11}-2(\epsilon_0+\epsilon_2)+h+\Lambda^2k^2+\frac{\Lambda^2 \pi^2}{t^2},\\
\Omega_\mathrm{1,z}&=h+P_\mathrm{11}+\left(k^2+\frac{\pi^2}{t^2}\right) \Lambda^2,
\end{split}
\end{equation*}
with $\Lambda=\sqrt{\frac{2 A}{\mu_\mathrm{0} M_\mathrm{s}^2 }}$ the exchange length, $h=H_\mathrm{0}/M_\mathrm{s}$, the dipole factors $P_\mathrm{00}=1-\frac{1-e^{-|k| t}}{|k| t}$, $P_\mathrm{11}=\frac{(k t)^2}{\pi^2+(kt)^2}\left(1-\frac{2(k t)^2}{\pi^2+(kt)^2}\frac{1+e^{-|k| t}}{|k| t} \right)$ and $Q=\frac{\sqrt{2}kt}{\pi^2+(kt)^2}\left(1+e^{-|k| t} \right)$. For a more detailed explanation of what all these factors describe, and the physical interpretation of this matrix, we refer the reader to Refs.~\onlinecite{PhysRevB.93.054430,doi:10.1063/1.4789962}. Lastly, we have the effective difference in anisotropy $\delta= \frac{2\sqrt{2}}{t\mu_\mathrm{0} M_\mathrm{s}^2}\left(\Delta K_\mathrm{s}-\frac{K_\mathrm{v,1} t~u(t-t_\mathrm{cr})\mathrm{sin}(\frac{\pi t_\mathrm{cr}}{t})}{\pi} \right) $, and sum anisotropies $\epsilon_\mathrm{0}=\frac{2}{t\mu_\mathrm{0} M_\mathrm{s}^2}\left(K_\mathrm{s,bot}+K_\mathrm{s,top} \right)$, $\epsilon_\mathrm{1}=\frac{2}{t\mu_\mathrm{0} M_\mathrm{s}^2}\left( K_\mathrm{v,0}t+K_\mathrm{v,1}(t-t_\mathrm{cr})u(t-t_\mathrm{cr}) \right)$ and \newline $\epsilon_\mathrm{2}=\frac{1}{t\mu_\mathrm{0} M_\mathrm{s}^2}\left( K_\mathrm{v,0}t+\frac{K_\mathrm{v,1}}{t 2\pi}u(t-t_\mathrm{cr})\left((-t+t_\mathrm{cr})2 \pi+t\mathrm{sin}(\frac{2\pi t_\mathrm{cr}}{t}) \right) \right)$. With respect to Ref.~\onlinecite{PhysRevB.93.054430}, note the sign difference of $\delta$, which we believe to be an typographical error. We can solve \verref{eq:Mat} by setting the total determinant to $0$, which results in the following equation
\begin{equation}
\label{eq:disp}
(\Omega_\mathrm{0}^2-\Omega^2)(\Omega_\mathrm{1}^2-\Omega^2)+2Q\delta\Omega(\Omega_\mathrm{0,z}-\Omega_\mathrm{1,z})-\delta^2 \Omega_\mathrm{0,z} \Omega_\mathrm{1,z}=0,
\end{equation}
with
\begin{equation*}
\Omega_\mathrm{0,1}^2=\frac{\Omega_\mathrm{00}^2+\Omega_\mathrm{11}^2  }{2}-Q^2 \mp \frac{1}{2}\sqrt{\left(\Omega_\mathrm{11}^2-\Omega_\mathrm{00}^2 \right)^2-4Q^2\left(\Omega_\mathrm{00}^2+\Omega_\mathrm{11}^2-\Omega_\mathrm{0,z}\Omega_\mathrm{1,x}-\Omega_\mathrm{0,x}\Omega_\mathrm{1,z} \right)}
\end{equation*}
which are the spin wave frequencies when $\delta=0$ and $\Omega_\mathrm{ii}=\sqrt{\Omega_\mathrm{i,x}\Omega_\mathrm{i,z}}$. These equations are once again slightly different with respect to Ref.~\onlinecite{PhysRevB.93.054430}. In~\verref{eq:disp} we have isolated a term linear in $Q$, which is the only term that changes sign when going from $k$ to $-k$ and is thus the only term that induce a frequency difference between counter-propagating waves. We can extract the frequency difference between those counter-propagating waves by assuming the frequency difference induced by a small $\delta$ leads to a minor modification to $\Omega_\mathrm{0}$. For this, we substitute into \verref{eq:disp} $\Omega=\Omega_\mathrm{0}+\delta\Omega$ and solve for $\delta\Omega$ to linear order in both $\delta\Omega$ and $\delta$ which gives as a frequency difference $\Delta f$ between two counter-propagating waves ($k<0$ - $k>0$):
\begin{equation}
\Delta f=\frac{\gamma\mu_\mathrm{0} M_\mathrm{s}}{2\pi }2 Q \delta \frac {\Omega_\mathrm{0,z}-\Omega_\mathrm{1,z}}{\Omega_\mathrm{0}^2-\Omega_\mathrm{1}^2 }.
\end{equation}
This equation is identical to the one provided in Ref.~\onlinecite{PhysRevB.93.054430} in the limit of $K_\mathrm{v}\rightarrow 0$ but is now modified by the additional volume terms in $\delta$ that can also lead to a shift. We combine this shift with the iDMI $D_\mathrm{s}$ induced frequency non-reciprocity~\cite{PhysRevB.88.184404}
\begin{equation}
\Delta f= \frac{2\gamma k D_\mathrm{s}}{\pi t M_\mathrm{s}}+\frac{\gamma\mu_\mathrm{0} M_\mathrm{s}}{2\pi }2 Q \delta \frac {\Omega_\mathrm{0,z}-\Omega_\mathrm{1,z}}{\Omega_\mathrm{0}^2-\Omega_\mathrm{1}^2 }
\end{equation}
for the complete frequency non-reciprocity.
\bibliography{../references,NMR_Co}